\documentclass[aps,prc,twocolumn,amsmath,amssymb,showpacs,nobibnotes,nofootinbib]{revtex4}
\usepackage{graphics}
\usepackage{mathptmx}
\usepackage{epsfig}

\begin{document}

\title{Nuclear level densities and $\gamma$-ray strength functions in $^{44,45}$Sc}

\author{A.C.~Larsen$^1$\footnote{E-mail: a.c.larsen@fys.uio.no}, 
M.~Guttormsen$^1$, R.~Chankova$^2$, F.~Ingebretsen$^1$, T.~L\"{o}nnroth$^3$, S.~Messelt$^1$, 
J.~Rekstad$^1$, A.~Schiller$^4$, S.~Siem$^1$, N.U.H.~Syed$^1$, and A.~Voinov$^5$
\\}

\affiliation{$^1$ Department of Physics, University of Oslo, P.O.Box 1048 Blindern, N-0316 Oslo, Norway\\
$^2$ North Carolina State University, Raleigh, NC 27695, USA \\
	and Triangle Universities Nuclear Laboratory, Durham, NC 27708 USA\\
$^3$ Department of Physics, \AA bo Akademi University, FIN-20500 \AA bo, Finland\\
$^4$ National Superconducting Cyclotron Laboratory, Michigan State University, East Lansing, MI 48824, USA\\
$^5$ Department of Physics and Astronomy, Ohio University, Athens, Ohio 45701, USA\\
and Frank Laboratory of Neutron Physics, Joint Institute for Nuclear Research,\\ 
141980 Dubna, Moscow region, Russia}

\date{\today}

\begin{abstract}
The scandium isotopes $^{44,45}$Sc have been studied with the $^{45}$Sc($^3$He,$\alpha \gamma$)$^{44}$Sc and $^{45}$Sc($^3$He,$^3$He$^{\prime} \gamma$)$^{45}$Sc reactions, respectively. The nuclear level densities and $\gamma$-ray strength functions have been extracted using the Oslo method. 
The experimental level densities are compared to calculated level densities obtained from a microscopic model based on BCS quasiparticles within the Nilsson level scheme. 
This model also gives information about the parity distribution and the number of broken Cooper pairs as a function of excitation energy. 
The experimental $\gamma$-ray strength functions are compared to theoretical models of the $E1$, $M1$, and $E2$ strength, and to data from ($\gamma$,n) and ($\gamma$,p) experiments. The strength functions show an enhancement at low $\gamma$ energies that cannot be explained by the present, standard models. 
\end{abstract} 


\maketitle

\section{Introduction}

The energy levels of an atomic nucleus and the decay probability of each level contain essential information on the nuclear structure. When the nucleus is excited to levels just above the ground state, spectroscopic measurements are able to give accurate information on the energy, spin, parity, and transition rates of the levels. However, as the excitation energy increases, the number of levels quickly becomes so high that all levels cannot be found with present state-of-the art spectroscopy methods. The nucleus leaves the discrete region and enters the region of quasi-continuum and continuum, where it is regarded as more appropriate to use average quantities to describe the behaviour of the nucleus. 

The nuclear level density and the $\gamma$-ray strength function give a measure of the gross properties of the nucleus. These average quantities are indispensable in practical applications of nuclear physics, such as calculations of nuclear reaction rates in astrophysical processes, the design and operation of fission reactors, and transmutation of nuclear waste. When it comes to fundamental nuclear structure, the level density can reveal information on, e.g., pair correlations and thermodynamic quantities such as entropy and temperature~\cite{melb0,gutt3}, while the average electromagnetic properties are characterized by the $\gamma$-ray strength function~\cite{blatt&weisskopf}. 

Neutron (and proton) resonance experiments provide data on the level density at or above the nucleon binding energy~\cite{egidy0}, and fluctuation analysis of total neutron cross sections~\cite{grimes} gives level density at excitation energies well above the nucleon binding energy. However, in the intermediate region between the nucleon binding energy and the discrete regime (the quasi-continuum), relatively little is known. To fill in this gap, the Oslo Cyclotron group has developed the so-called Oslo method, which enables the extraction of both level density and $\gamma$-ray strength function from the distribution of primary $\gamma$ rays at various initial excitation energies. The method has been thoroughly tested on nuclei in the rare-earth region~\cite{sunniva,bagheri,undraa}, and has also been successfully extended to other mass regions~\cite{Si,Fe&Mo,chankova,larsen}. 

The present work reports on new results from an experiment on the scandium isotopes $^{44,45}$Sc. The $^{45}$Sc nucleus has one unpaired proton in the $\pi f_{7/2}$ orbital, while $^{44}$Sc has an unpaired proton and a neutron in the $\pi f_{7/2},\nu f_{7/2}$ orbitals. If one naively assumes that only the $f_{7/2}$ orbital is dominant in the model space, one would expect a majority of positive parity states in the case of $^{44}$Sc, and negative parity states for $^{45}$Sc. However, it is well known that states with different parity appear already at very low excitation energy in these nuclei. Early attempts on reproducing the states both with particle-plus-rotor models~\cite{Malik&Scholz} and  shell-model calculations~\cite{cole} had relatively little success. More recent works have shown that these nuclei exhibit both collective and single-particle character even at low excitation energy, and they have been considered as a good case for studying the interplay between the single-particle and the collective degrees of freedom in medium-mass nuclei near the closed shell~\cite{broda,caurier}. These scandium isotopes are therefore of special interest to test the Oslo method further.

In Sect.~II an outline of the experimental procedure and the Oslo method is given. The level densities and the  $\gamma$-ray strength functions are discussed in Sects.~III and IV, respectively. 
Finally, concluding remarks are given in Sect.~V.

\section{Experimental details and the Oslo Method}

The experiment was performed at the Oslo Cyclotron Laboratory (OCL) using a beam of $^3$He-ions with energy 38 MeV. The self-supporting natural target of $99.9$\% $^{45}$Sc had a thickness of 3.4 mg/cm$^2$. Eight Si $\Delta E - E$ telescopes were arranged close to the target at an angle of 45$^{\circ}$ relative to the beam. The $\gamma$-detector array CACTUS~\cite{Cactus}, consisting of 28 collimated NaI crystals with a total efficiency of $\sim$15\%, surrounded the target and the particle detectors. The experimental setup enabled particle-$\gamma$ coincidence measurements of the reactions ($^3$He,$\alpha \gamma$) and ($^3$He,$^3$He$^{\prime}\gamma$). These reactions populate states with spin range $I\sim 2-6\hbar$, which means that most of the energy transferred to the target nucleus is intrinsic excitation energy. The experiment ran for about five days, with a typical beam current of $\sim 1$ nA.

\begin{figure}[b]
\centering
\includegraphics[height=15cm]{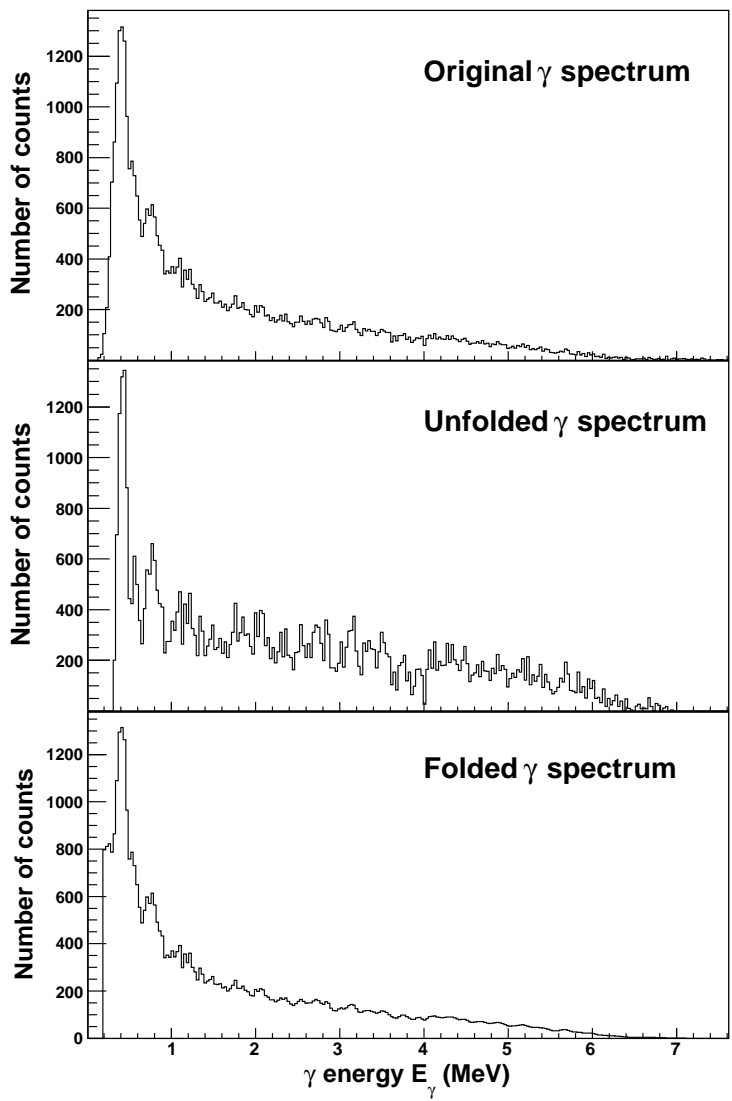}
\caption{Original (top), unfolded (middle) and folded $\gamma$ spectrum of $^{44}$Sc for excitation energy between $5.5-6.5$ MeV.}
\label{fig:unfold}
\end{figure}

\begin{figure}[b]
\centering
\includegraphics[height=15cm]{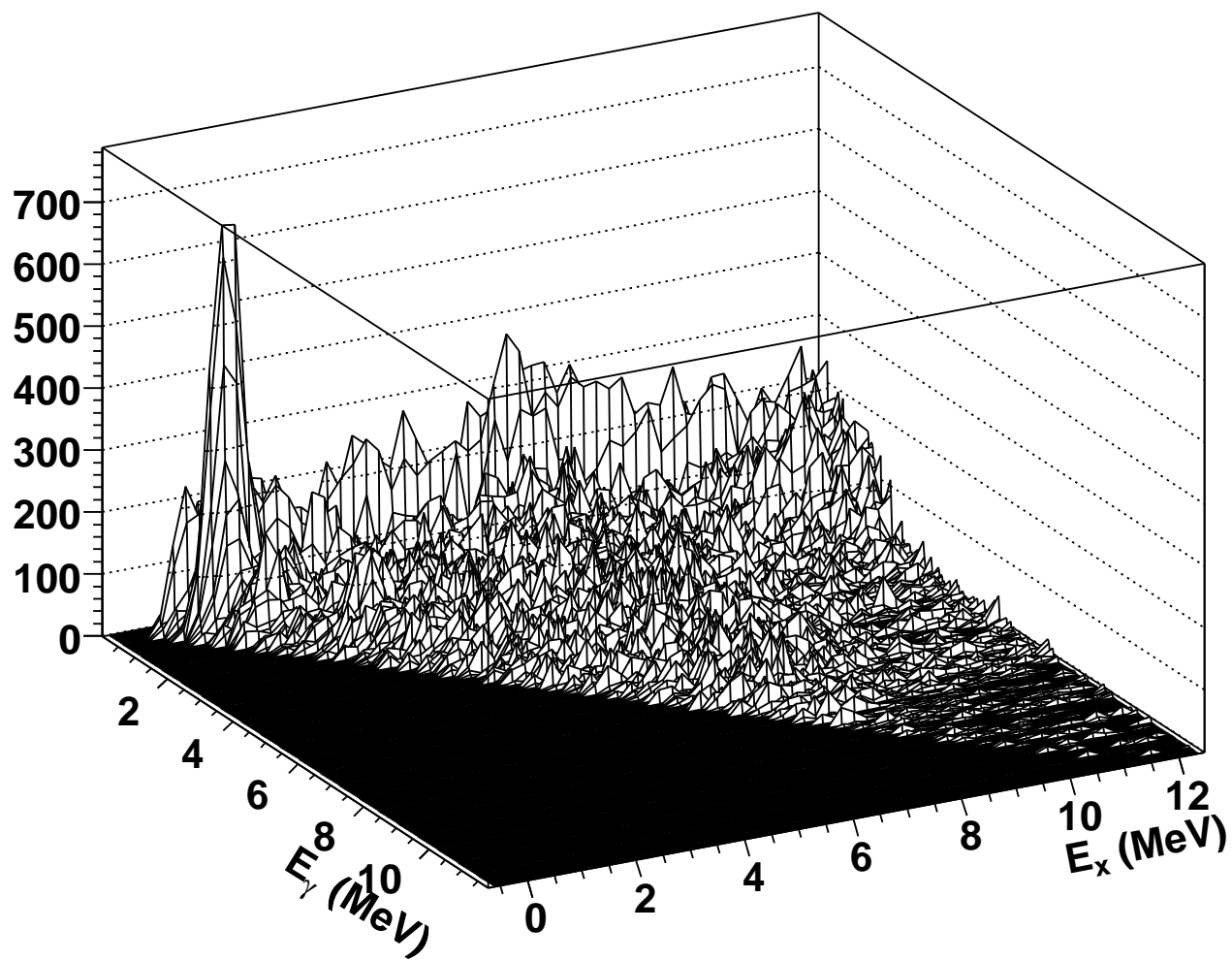}
\caption{Unfolded particle-$\gamma$ matrix for the $^{45}$Sc($^{3}$He,$\alpha$)$^{44}$Sc reaction.}
\label{fig:alfna}
\end{figure}

The recorded coincidences were sorted into two-dimensional particle-$\gamma$ matrices. From the reaction kinematics, the measured energy of the outgoing $^3$He or $\alpha$ particle were converted into excitation energy of the residual nucleus. With particle-energy bins of 240 keV/channel, total $\gamma$-ray spectra were obtained for each bin. These $\gamma$ spectra were then unfolded using a well-tested unfolding procedure based on the known response functions of the CACTUS array~\cite{gutt6}. The unfolding method described in Ref.~\cite{gutt6} preserves the fluctuations in the original spectra without introducing further, spurious fluctuations. In Fig.~\ref{fig:unfold} an original $\gamma$ spectrum, an unfolded spectrum, and the unfolded spectrum convoluted with the response functions are shown for $^{44}$Sc with gate on the excitation-energy bins between $5.5 - 6.5$ MeV. The original and the convoluted spectrum show excellent agreement, giving strong confidence in the unfolding method. The unfolded particle-$\gamma$ matrix of the $^{45}$Sc($^3$He,$\alpha \gamma$)$^{44}$Sc data is displayed in Fig.~\ref{fig:alfna}, where the sharp diagonal $E = E_{\gamma}$ is clearly seen. Apart from the prominent peak at $E\sim 1$ MeV and $E_{\gamma} \sim 0.75$ MeV, the matrix is without outstanding structures. 

\begin{figure}[htb]
\centering
\includegraphics[height=15cm]{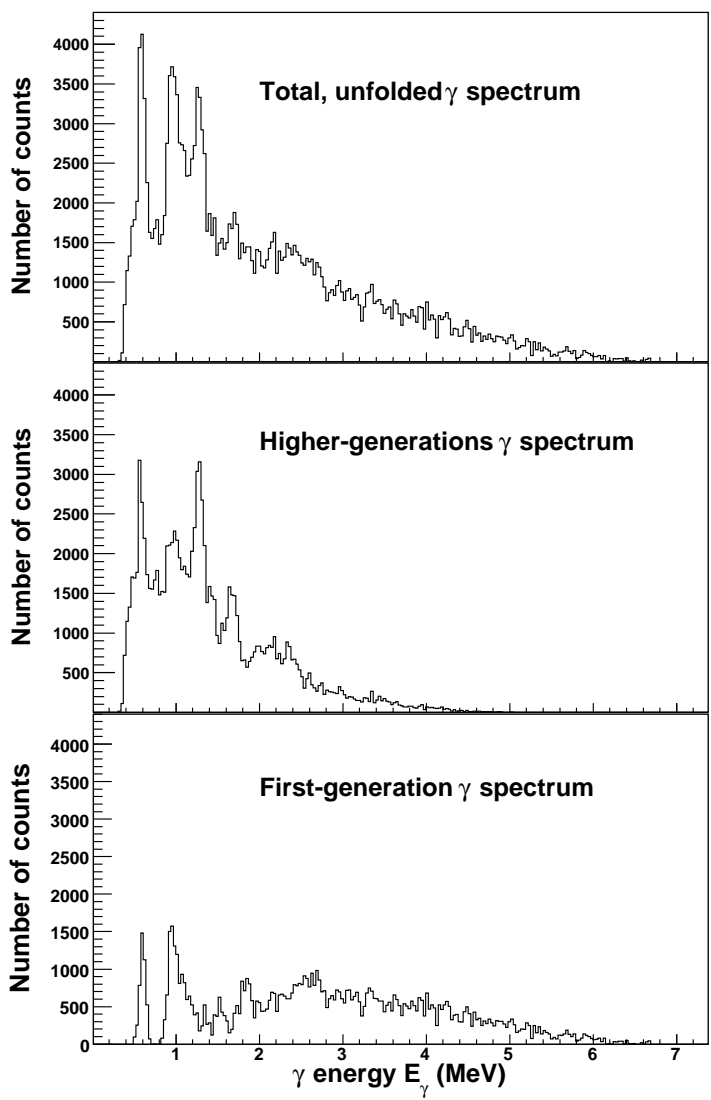}
\caption{Unfolded, total $\gamma$ spectrum, second and higher-generation $\gamma$ spectrum and first-generation $\gamma$ spectrum of $^{45}$Sc for excitation energy between $5.5-6.5$ MeV.}
\label{fig:firstgen}
\end{figure}

The energy distribution of the first emitted $\gamma$ rays from the decay cascades reveals essential information on the nuclear structure. In order to extract these primary $\gamma$ rays from the total $\gamma$ spectra, a subtraction procedure described in Ref.~\cite{gutt0} is applied for each excitation-energy bin. The main assumption of this method is that the $\gamma$ decay from any excitation-energy bin is independent on how the nucleus was excited to this bin. In other words, the decay routes are the same whether they were initiated directly by the nuclear reaction or by $\gamma$ decay from higher-lying states. This assumption is automatically fulfilled when the same states are equally populated by the two processes, since $\gamma$ branching ratios are properties of the levels themselves. Even if different states are populated, the assumption is still valid for statistical $\gamma$ decay, which only depends on the $\gamma$-ray energy and the number of accessible final states. Figure~\ref{fig:firstgen} shows the total, unfolded $\gamma$ spectrum, the second and higher generations $\gamma$ spectrum and the first-generation spectrum of $^{45}$Sc for excitation energy between $E = 5.5 - 6.5$ MeV. The first-generation spectrum is obtained by subtracting the higher-generation $\gamma$ rays from the total $\gamma$ spectrum. By looking at the lower panel of Fig.~\ref{fig:firstgen}, it is clear that the main assumption of the subtraction method is not fulfilled for $E_{\gamma} \lesssim 1.4$ MeV. In this region, some strong, low-energy transitions were not subtracted correctly. This means that the levels from which these transitions originate are populated more strongly from higher excited levels through $\gamma$ emission, than directly by inelastic $^{3}$He scattering. Therefore, only data for $E_{\gamma} > 1.6$ MeV is used in the further analysis. Similar considerations were done for $^{44}$Sc. 

The experimental matrix of first-generation $\gamma$ rays is then normalized~\cite{schi0} such that for every excitation-energy bin $E$, the sum over all $\gamma$ energies $E_{\gamma}$ from some minimum value  $E_{\gamma}^{min}$ to the maximum value $E_{\gamma}^{max}=E$ at this excitation-energy bin is unity:
\begin{equation}
\sum_{E_{\gamma}=E_{\gamma}^{min}}^{E} P(E, E_{\gamma}) = 1.
\label{eq:matrixnorm}
\end{equation}
For statistical $\gamma$ decay in the continuum region, the $\gamma$-decay probability from an excitation energy $E$ to $E_{f}= E-E_{\gamma}$ is proportional to the $\gamma$-ray transmission coefficient ${\mathcal T}(E_{\gamma})$ and the level density at the final excitation energy $\rho (E_{f})$:
\begin{equation}
P(E, E_{\gamma}) \propto  \rho (E -E_{\gamma}) {\mathcal{T}}  (E_{\gamma}).
\label{eq:axel}
\end{equation}
The essential assumption underlying the above relation is that the reaction can be described as a two-stage process, where a compound state is first formed, before it decays in a manner that is independent of the mode of formation~\cite{BM,hend1}. Equation~(\ref{eq:axel}) could also be regarded as a generalization\footnote{A generalization in the sense that the present work deals with an ensemble of initial and final states, and therefore considers the average decay properties in each excitation-energy bin.} of Fermi's golden rule, where the decay rate is proportional to the density of final states and the square of the matrix element between the initial state and the final state. 

The experimental normalized first-generation $\gamma$ matrix can theoretically be approximated by 
\begin{equation}
P_{th}(E, E_{\gamma}) = \frac {\rho (E -E_{\gamma}) {\mathcal{T}}  (E_{\gamma})}{\sum_{E_{\gamma}=E_{\gamma}^{min}}^{E} \rho (E -E_{\gamma}) {\mathcal{T}}  (E_{\gamma})}.
\label{eq:theory}
\end{equation}
The $\gamma$-ray transmission coefficient ${\mathcal T}$ is independent of excitation energy according to the generalized Brink-Axel hypothesis~\cite{brink,axel}, which states that collective excitation modes built on excited states have the same properties as those built on the ground state. There is evidence that the width of the giant dipole resonance (GDR) varies with the nuclear temperature of the state on which it is built~\cite{kad,Ger}. However, the temperature corresponding to the excitation-energy range covered in this work is rather low and changes slowly with excitation energy ($T \sim \sqrt{E_{f}}$ ). The temperature is therefore assumed to be approximately constant, and the Brink-Axel hypothesis is recovered in the energy region of interest. 

\begin{figure}[h]
\centering
\includegraphics[height=15cm]{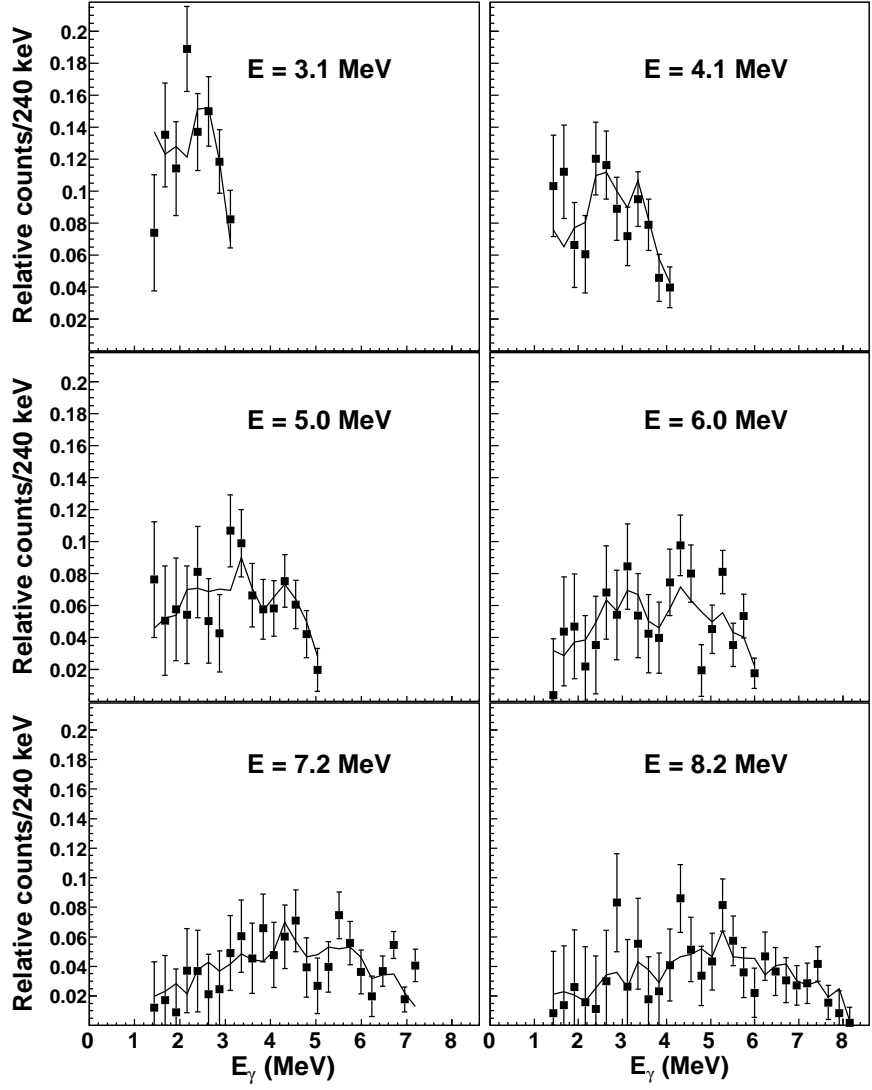}
\caption{A sample of the experimental first-generation spectra of $^{44}$Sc (data points with error bars) are plotted with the least-$\chi^{2}$ fit (lines).}
\label{fig:work}
\end{figure}

To extract the level density and the $\gamma$-ray transmission coefficient, an iterative procedure~\cite{schi0} is applied to the first-generation $\gamma$ matrix $P(E, E_{\gamma})$. The basic idea of this method is to minimize
\begin{equation}
\chi^{2} = \frac{1}{N_{free}}\sum_{E=E^{min}}^{E^{max}}\sum_{E_{\gamma}=E_{\gamma}^{min}}^{E} \left( \frac{P_{th}(E, E_{\gamma}) - P(E, E_{\gamma})}{\Delta P(E, E_{\gamma})} \right)^{2},
\label{eq:chisquared}
\end{equation}
where $N_{free}$ is the number of degrees of freedom, and $\Delta P(E, E_{\gamma})$ is the uncertainty in the experimental first-generation $\gamma$ matrix. Every point of the $\rho$ and ${\mathcal{T}}$ functions is assumed as an independent variable, so the reduced $\chi^{2}$ is minimized for every argument $E-E_{\gamma}$ and $E$. The quality of the procedure when applied to the $^{44}$Sc data is shown in Fig.~\ref{fig:work}, where the experimental first-generation spectra for various initial excitation energies are compared to the least-$\chi^{2}$ solution. In general, the agreement between the experimental data and the fit is very good. 

The globalized fitting to the data points only gives the functional form of $\rho$ and ${\mathcal{T}}$. In fact, it has been shown \cite{schi0} that if one solution for the multiplicative functions $\rho$ and ${\mathcal{T}} $ is known, one may construct an infinite number of other functions, which give identical fits to the $P(E, E_{\gamma})$ matrix by
\begin{eqnarray}
\tilde{\rho}(E-E_\gamma)&=&A\exp[\alpha(E-E_\gamma)]\,\rho(E-E_\gamma),
\label{eq:array1}\\
\tilde{{\mathcal{T}}}(E_\gamma)&=&B\exp(\alpha E_\gamma){\mathcal{T}} (E_\gamma).
\label{eq:array2}
\end{eqnarray}
Therefore the transformation parameters $\alpha$, $A$ and $B$, which correspond to the physical solution, remain to be found.

\section{The level densities}

\subsection{Normalization}

As described in the previous section, only the shape of the level density is found through the least $\chi^{2}$ procedure of~\cite{schi0}. To determine the slope $\alpha$ and the absolute value $A$ in Eq.~(\ref{eq:array1}), the $\rho$ function is adjusted to match the number of known discrete levels at low excitation energy~\cite{ENSDF} and proton-resonance data~\cite{poirier,mitchell} at high excitation energy. The procedure for extracting the total level density $\rho$ from the resonance spacing $D$ is described in Ref.~\cite{schi0}. Since the proton beam energy had a range of $E_{p}$($^{44}$Sc)$ = 0.90 - 1.50$ MeV and $E_{p}$($^{45}$Sc)$ = 2.50 - 3.53$ MeV in \cite{poirier} and \cite{mitchell} respectively, the level density estimated from the proton resonances is not at the proton binding energy $B_p$, but rather at $\sim B_{p}+(\Delta E)/2$, where $\Delta E$ is the energy range of the proton beam, assuming that the resonances are approximately equally distributed over $\Delta E$. Also, the authors of~\cite{poirier} do not distinguish between s- and p-wave resonances, so the calculation of the total level density is rather uncertain in the case of  $^{44}$Sc. However, by comparing with preliminary level-density data from an experiment done on $^{44}$Sc at Ohio University, the slope $\alpha$ seems to be correct~\cite{private}.

\begin{figure}[htb]
\centering
\includegraphics[height=15cm]{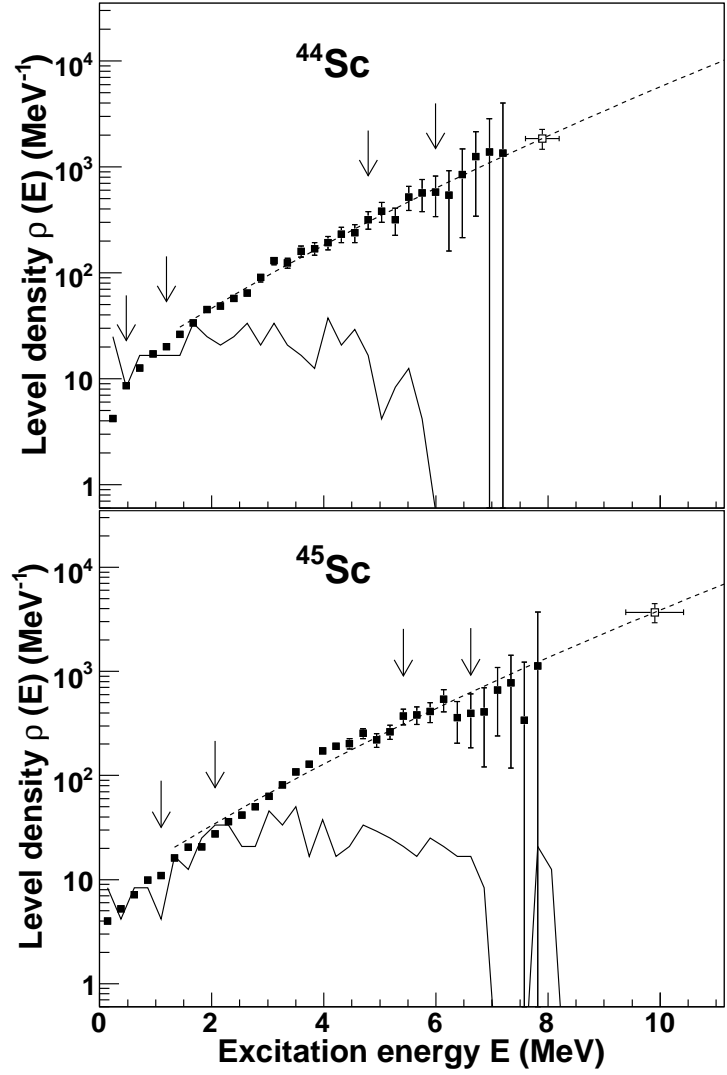}
\caption{Normalization procedure of the experimental level density (data points) of $^{44,45}$Sc. The data points between the arrows are normalized to known levels at low excitation energy (solid line) and to the level density at the proton-separation energy (open square) using the Fermi-gas level density (dashed line).}
\label{fig:counting}
\end{figure}

Because our experimental data points of the level density only reach up to an excitation energy of $\sim $7.2 and $\sim $8.0 MeV for $^{44,45}$Sc respectively, we extrapolate with the back-shifted Fermi gas model~\cite{GC,egidy} 
\begin{equation} 
\rho_{\rm BS}(E)= \eta\frac{\exp(2 \sqrt{aU})}{12 \sqrt{2}a^{1/4}U^{5/4} \sigma}, 
\label{eq:bs}
\end{equation}
where a constant $\eta$ is introduced to ensure that $\rho_{\rm BS}$ has the same value as the level density calculated from the proton-resonance experiments. The intrinsic excitation energy is estimated by $U=E-E_1$, where $E_1$ is the back-shift parameter. The spin-cutoff parameter is given by\footnote{The authors of~\cite{egidy} found this expression to be the most adequate in the low-energy region, even though it is connected to the (mathematically incorrect) relation $U = aT^{2} - T$, and not the standard one $U = aT^{2}$ (See Ref.~\cite{GC} for more details).}
\begin{equation}
\sigma^{2}  =  0.0146A^{5/3} \frac{1+\sqrt{1+4aU}}{2a},
\label{eq:sigma}
\end{equation} 
where $A$ is the mass number. Since the level density parameter $a$ and the back-shift parameter $E_1$ calculated with the method of Ref.~\cite{egidy} did not seem to give reliable results for $^{45}$Sc, these parameters were extracted by fitting the Fermi gas to the known levels at $\sim 1.75$ MeV and $\sim 2$ MeV for $^{44,45}$Sc, respectively, and to the known resonance-spacing data at $B_{p}+(\Delta E)/2$. The parameters used for $^{44,45}$Sc in Eq.~(\ref{eq:bs}) are listed in Table~\ref{tab:tab1}, where also the Fermi-gas parameters from ~\cite{egidy} are shown. As the authors demonstrate in Fig.~5 in Ref.~\cite{egidy}, the difference between the calculated parameters and the empirically extracted ones might be large in the mass region $A \leq 50$. The normalization procedure is pictured in Fig.~\ref{fig:counting}; note that only statistical errors are shown. Above $\sim 2$ MeV, there are more than 30 levels per MeV, giving the present limit to make complete spectroscopy in these nuclei. 
 
\begin{figure}[htb]
\centering
\includegraphics[height=15cm]{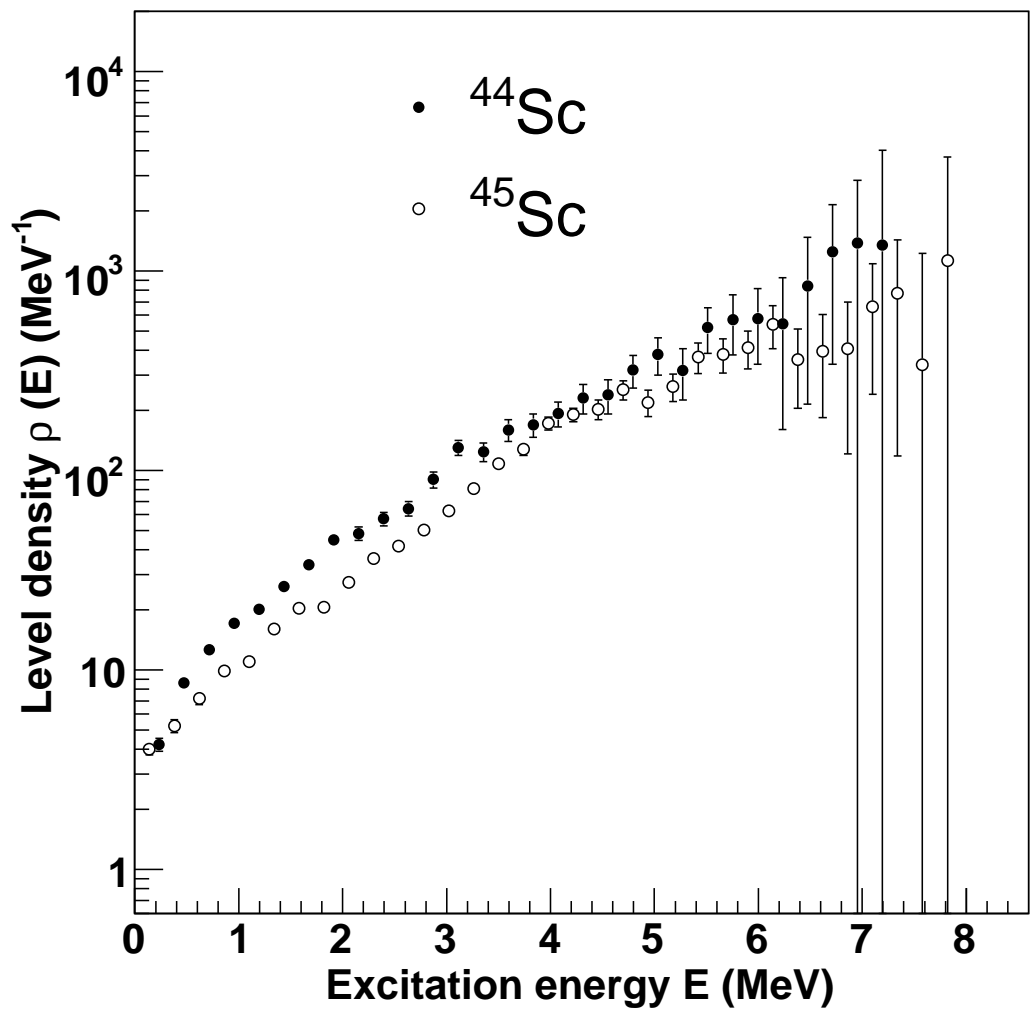}
\caption{Normalized level densities for $^{44,45}$Sc.}
\label{fig:rhoboth}
\end{figure}

The normalized level densities of $^{44}$Sc and $^{45}$Sc are displayed in Fig.~\ref{fig:rhoboth}. As one would expect, the odd-odd nucleus $^{44}$Sc has an overall higher level density than its odd-even neighbour $^{45}$Sc due to its two unpaired nucleons. The difference in level density between the odd-odd ($^{44}$Sc) and the odd-even ($^{45}$Sc) nucleus is seen to be approximately constant, except in the area between $E \sim 4 - 5$ MeV, where the level densities are almost the same. This is in agreement with earlier findings in the rare-earth region. However, here the odd-odd system has approximately a factor of two higher level density compared to the odd-even nucleus, while for rare-earth nuclei the difference was found to be a factor of five. 

Bump structures in the level densities of the scandium nuclei are observed. Standard models such as the back-shifted Fermi gas give a smooth $\rho$ function, and are unable to describe the structures that appear in the experimental level density in this excitation-energy region. 

\subsection{Comparison with microscopic model}

In order to further investigate the level density at high excitation energy, a microscopic model has been developed. The model is based on combining all possible proton and neutron configurations within the Nilsson energy scheme, and the concept of Bardeen-Cooper-Schrieffer (BCS) quasiparticles~\cite{BCS} is utilized. 

The model is described within the microcanonical ensemble, where the excitation energy $E$ is well defined. The single-particle energies $e_{\rm sp}$ are taken from the Nilsson model for an axially deformed core described by the quadrupole deformation parameter $\epsilon_2$. Furthermore, the model depends on the spin-orbit and centrifugal parameters $\kappa$ and $\mu$. The oscillator quantum energy $\hbar \omega_0 = 41 A^{-1/3}$ MeV between the harmonic oscillator shells is also input to the code.
Within the BCS model, the single-quasiparticle energies are defined by
\begin{equation}
e_{\rm qp}=\sqrt{(e_{\rm sp} - \lambda)^2 +\Delta^2 },
\end{equation}
where the Fermi level $\lambda$ is adjusted to reproduce the number of particles in the system and $\Delta$ is the pair-gap parameter, which is kept constant.

The double-degenerated proton and neutron quasiparticle orbitals are characterized by their spin projections on the symmetry axis $\Omega_{\pi}$ and $\Omega_{\nu}$, respectively. The energy due to quasiparticle excitations is given by
\begin{equation}
E_{\rm qp}(\Omega_{\pi},\Omega_{\nu})=\sum_{ \left\{ \Omega_{\pi}^{\prime}\Omega_{\nu}^{\prime} \right\} }
 \left[ e_{\rm qp}(\Omega_{\pi}^{\prime})+e_{\rm qp}(\Omega_{\nu}^{\prime})
  + V(\Omega_{\pi}^{\prime},\Omega_{\nu}^{\prime}) \right].
\label{eq:sum}
\end{equation}
Between the aligned and anti-aligned levels of the proton and neutron projections, i.e. $\Omega_{\pi} + \Omega_{\nu}$ and $|\Omega_{\pi} - \Omega_{\nu}|$, a residual interaction $V$ is defined as a random Gaussian distribution centered at zero energy with a width of 50 keV. The sets of proton and neutron orbitals $\left\{ \Omega_{\pi}^{\prime}\Omega_{\nu}^{\prime} \right\}$ are picked out by using a random generator. The total number of broken Cooper pairs are set to three, making a maximum number of eight participating quasi-particles for odd-odd nuclear systems. Technically, this process is repeated until all possible energies $E_{\rm qp}(\Omega_{\pi},\Omega_{\nu})$ have been obtained. An indicator that this saturation is reached, is that all energies are reproduced at least ten times in the simulation.

Collective energy terms are schematically added by
\begin{equation}
E = E_{\rm qp}(\Omega_{\pi},\Omega_{\nu})  
+ A_{\rm rot}R(R+1) + \hbar\omega_{\rm vib}\nu,
\label{eq:tot}
\end{equation}
where $A_{\rm rot}= \hbar^{2}/2{\mathcal J}$ is the rotational parameter and $R=0,1,2,3\ldots$ is the rotational quantum number. The vibrational motion is described by the phonon number $\nu = 0,1,2,\ldots$ and the oscillator quantum energy $\hbar\omega_{\rm vib}$.

The advantage of the present model is a fast algorithm that may include a large model space of single-particle states. Since level density is a gross property, the detailed knowledge of the many-particle matrix elements through large diagonalizing algorithms is not necessary. No level inversion is observed, as frequently seen for microscopic models with single-particle orbital truncations. In the sum of Eq.~(\ref{eq:sum}), all orbitals with energy up to the maximum energy ($e_{\rm qp} < E$) are included. Typically, for excitation energies up to $\sim 10$ MeV, about 20 proton and 20 neutron orbitals are taken into account ($\sim 10$ orbitals below the Fermi level and $\sim 10$ orbitals above).  

\begin{figure}[htb]
\centering
\includegraphics[height=15cm]{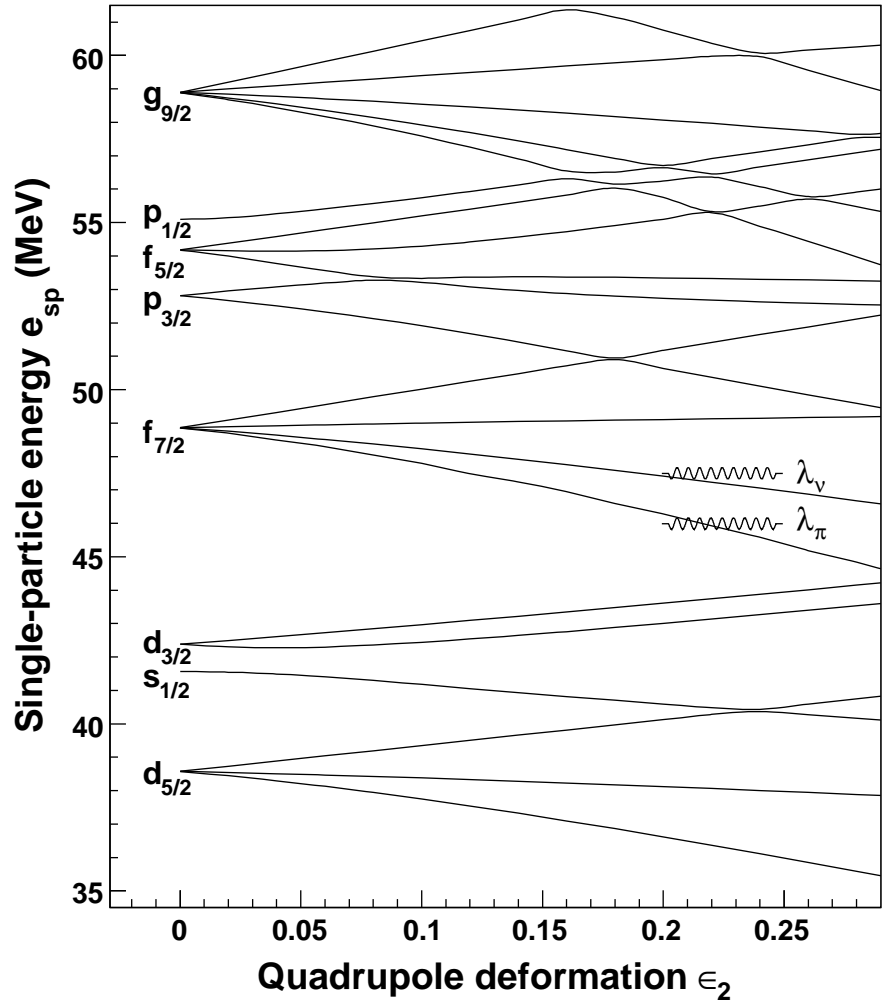}
\caption{The Nilsson level scheme for $^{45}$Sc with parameters $\kappa = 0.066$ and $\mu = 0.32$.}
\label{fig:nilsson}
\end{figure}

In the calculation we have adopted the Nilsson parameters $\kappa=0.066$ and $\mu=0.32$ from~\cite{white} with oscillator quantum energy of $\hbar \omega_{\rm vib} = 1.904$ MeV, found from the $0^{+}$ vibrational state in $^{44}$Ti \cite{ToI}. The Nilsson levels used in the calculations for $^{45}$Sc are shown in Fig.~\ref{fig:nilsson}, with the Fermi levels for the protons and neutrons. The value of the deformation parameter $\epsilon_2$  was set to $0.23$, which is in agreement with values suggested in Ref.~\cite{broda}. The rotational and vibrational terms contribute only significantly to the total level density in the lower excitation region. To reproduce the transition energy from the $11/2^{-} \rightarrow 7/2^{-}$ transition in the ground-state rotational band of $^{45}$Sc~\cite{ToI}, the rotational parameter $A_{\rm rot}$ was set to $0.135$ MeV. The adopted pairing gap parameters $\Delta_{\pi}$ and $\Delta_{\nu}$ are taken from the calculations of Dobaczewski {\it et al.}~\cite{doba} for the even-even $^{42}$Ca for $^{44}$Sc and $^{44}$Ca for $^{45}$Sc. A list of the input data for the model calculations can be found in Table~\ref{tab:tab2}.

\begin{figure}[htb]
\centering
\includegraphics[height=15cm]{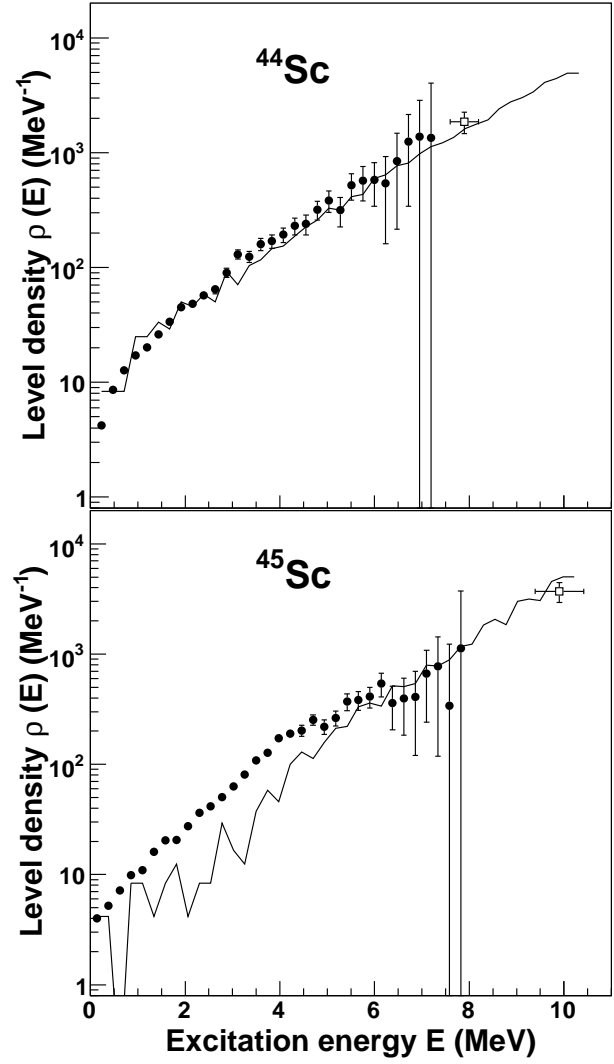}
\caption{Calculated level densities (solid lines) compared with the experimental ones (data points with error bars) for $^{44,45}$Sc.}
\label{fig:micro}
\end{figure}

The experimental and calculated level densities are shown in Fig.~\ref{fig:micro}. The result is satisfactory, especially for the nucleus $^{44}$Sc where there is a good agreement between the model calculation and the experimental level density. The general decrease in level density for the odd-even system compared to the odd-odd nucleus as well as the level densities found from the proton-resonance experiments are well reproduced. However, it is seen that the model misses many low-lying levels in the excitation-energy region $E=1-5$ MeV for $^{45}$Sc. This can, at least partially, be explained by the well-established shape coexistence determined from the negative-parity and positive-parity bands in this nucleus~\cite{broda}. Only one shape is included in our model, and thus only one potential, which results in an undershoot of bandheads of about a factor of two. 

The pairing parameters $\Delta_{\pi}$ and $\Delta_{\nu}$ are important inputs of the model, since the slope of the level density (in log scale) increases with decreasing pairing parameters in the energy region considered here. It can be seen from Fig.~\ref{fig:micro} that the adopted values give a nice agreement of the log slope of the level densities for both isotopes. 

\begin{figure}[htb]
\centering
\includegraphics[height=15cm]{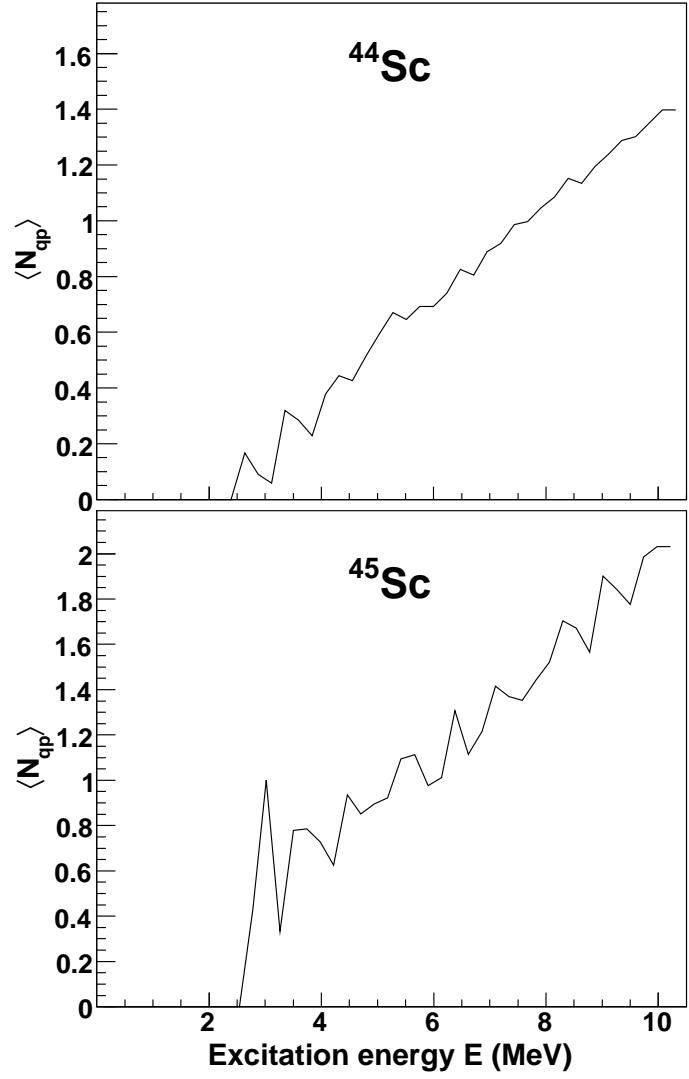}
\caption{The average number of broken Cooper pairs as function of excitation energy for $^{44,45}$Sc.}
\label{fig:pairs}
\end{figure}

Figure~\ref{fig:pairs} shows the average number of broken Cooper pairs $\langle N_{qp}\rangle$ as a function of excitation energy. This is calculated by looking at all configurations obtained in each 240-keV excitation-energy bin, and finding the number of configurations with one broken pair, two broken pairs and so on. Both neutron and proton pairs are taken into account. From this information the average number of broken Cooper pairs is calculated. From Fig.~\ref{fig:pairs}, the pair-breaking process is seen to start at $E\sim 2.5$ MeV for both nuclei, in accordance with the values used for $\Delta_{\pi}$ (see Table~\ref{tab:tab2}). The average number of broken pairs seems to have a relatively linear increase, giving an exponential growth in the level density. This behaviour also indicates that there is no abrupt change in seniority as a function of excitation energy. For example, in the region $E = 9-10$ MeV, the model predicts 1\% states with no pairs broken, 34\% states with one broken pair, 61\% states with two broken pairs, and 4\% of the states have three pairs broken. 

\begin{figure}[htb]
\centering
\includegraphics[height=15cm]{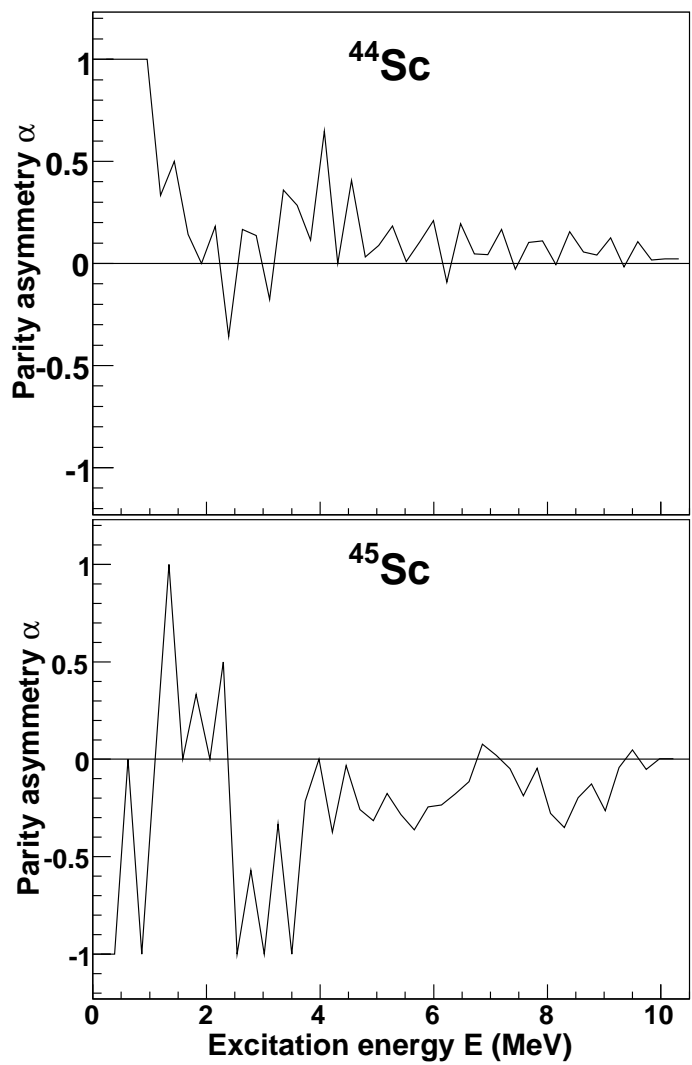}
\caption{The parity asymmetry as function of excitation energy for $^{44,45}$Sc.}
\label{fig:asymmetry}
\end{figure}

The location of the proton and neutron Fermi levels of $^{44,45}$Sc in the Nilsson level scheme gives, roughly speaking, mostly positive-parity orbitals below and negative-parity states above the Fermi levels. Knowing this, one would expect a relatively homogeneous mixture of positive and negative parity states in the whole excitation-energy region covered by the calculations. In order to investigate this feature, we utilize the parity asymmetry defined in Ref.~\cite{gary} by
\begin{equation}
\alpha=\frac{\rho_+-\rho_-}{\rho_++\rho_-},
\end{equation}
which gives $-1$ and $1$ for only negative and positive parities, respectively, and 0 when both parities are equally represented. In Fig.~\ref{fig:asymmetry} the parity asymmetry $\alpha$  is shown as a function of excitation energy. On the average, for $E > 4$ MeV, there seems to be a slight excess of positive and negative parity states in $^{44}$Sc and $^{45}$Sc, respectively. However, as the excitation energy increases, the model predicts that the parity asymmetry becomes smaller and smaller for both nuclei. The proton-resonance data in Ref.~\cite{gary} from the reaction $^{44}$Ca$+p$ (compound nucleus $^{45}$Sc, with excitation-energy region $9.77-10.53$ MeV), gives an asymmetry parameter $\alpha = -0.18^{+0.07}_{-0.06}$ for $J=1/2$ resonances, and $\alpha = 0.23\pm 0.07$ for $J=3/2$ resonances. Given the level densities of $J=1/2$ and $J=3/2$ resonances (see Table III in ~\cite{gary}), the parity asymmetry for $\rho(J=1/2,J=3/2)$ can be estimated to $\alpha \sim 0.02$, in good agreement with the model's result in this excitation-energy region.

\section{The $\gamma$-ray strength functions}

As mentioned in Sect.~II, the $\gamma$ decay process in the (quasi) continuum is governed by the level density and the $\gamma$-ray transmission coefficient. By using the Oslo method, also the $\gamma$-ray transmission coefficient can be extracted from the experimental data.

The slope of the $\gamma$-ray transmission coefficient ${\mathcal{T}} (E_{\gamma})$ has already been determined through the normalization of the level densities (Sect.~IIIA). However, the constant $B$ in Eq.~(\ref{eq:array2}) remains to be determined.  If there was data on the average total radiative width $\langle\Gamma_{\gamma} \rangle$ for these nuclei, this data could be utilized for the absolute normalization of ${\mathcal{T}}$ as described in, e.g,~\cite{voin1,gutt7}. Since such data does not exist for $^{44,45}$Sc, other considerations had to be made to obtain the absolute value of the strength function.

\begin{figure}[htb]
\centering
\includegraphics[height=15cm]{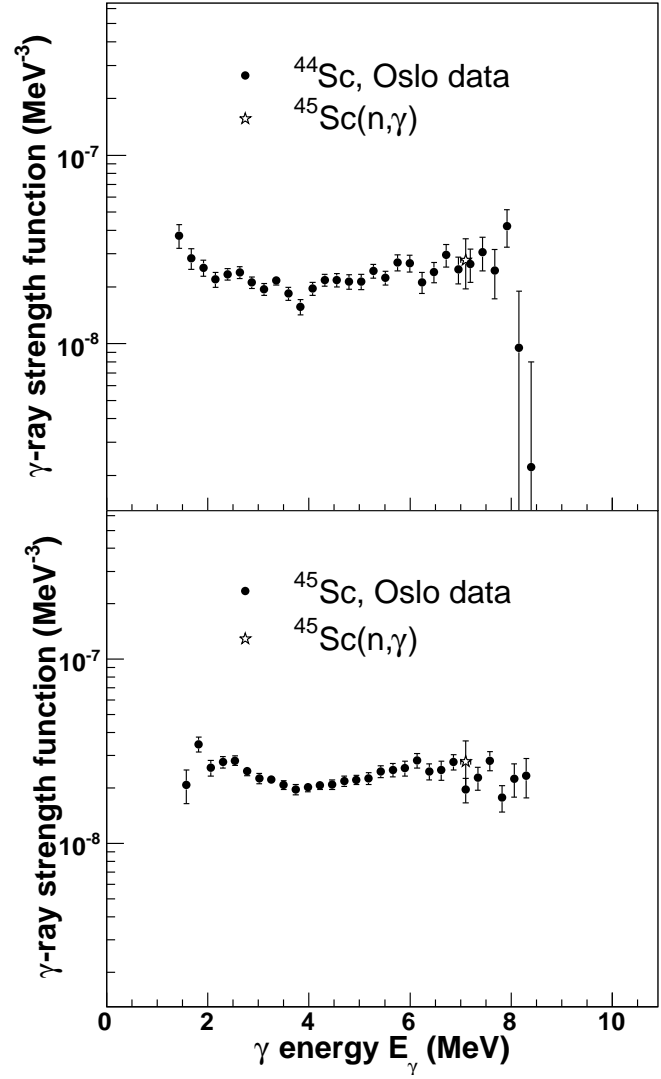}
\caption{Normalized $\gamma$-strength functions of $^{44,45}$Sc (black dots), and $f_{E1}+f_{M1}$ from Ref.~\cite{kopecky&uhl} (star).}
\label{fig:strengthboth}
\end{figure}

The experimental ${\mathcal{T}}$ contains components from all electromagnetic characters $X$ and multipolarities $L$. It is closely connected to the total $\gamma$-ray strength function through the relation~\cite{kopecky}
\begin{equation}
{\mathcal{T}}(E_{\gamma}) = 2\pi \sum_{XL} E_{\gamma}^{2L+1} f_{XL}(E_{\gamma}),
\end{equation}
where $f_{XL}$ is the $\gamma$-ray strength function for electromagnetic character $X$ and multipolarity $L$. Assuming that the $\gamma$-decay taking place in the continuum is dominated by $E1$ and $M1$ transitions, the total $\gamma$-ray strength function can be approximated by 
\begin{equation}
f(E_{\gamma}) \simeq \frac{1}{2\pi}\frac{ {\mathcal{T}} (E_{\gamma}) }{ E_{\gamma}^{3} }.
\end{equation}
The resulting $\gamma$-ray strength functions of $^{44,45}$Sc are then scaled to agree with data from Ref.~\cite{kopecky&uhl}. Based on two resonances from the reaction $^{45}$Sc(n,$\gamma$) and on the observation of 13 $E1$ transitions and 9 $M1$ transitions of average energy 7.0 and 7.2 MeV, respectively, the strength functions are found to be $f_{E1} = 1.61(59) \cdot 10^{-8}$ MeV$^{-3}$ and $f_{M1} = 1.17(59) \cdot 10^{-8}$ MeV$^{-3}$~\cite{kopecky&uhl}. By adding these values together, the absolute normalization is given at this specific $\gamma$ energy. The experimental $\gamma$-ray strength functions of $^{44,45}$Sc are displayed in Fig.~\ref{fig:strengthboth}, together with the data point from Ref.~\cite{kopecky&uhl} used for the normalization. 

Several interesting features can be seen in Fig.~\ref{fig:strengthboth}. In general, for $E_{\gamma} \geq 3.5$ MeV, the data show that the $\gamma$-ray strength functions of $^{44,45}$Sc are slowly increasing with $\gamma$ energy. For $\gamma$ energies below $\sim 3$ MeV, the $\gamma$-ray strength functions of both nuclei have an increase of a factor $\sim 3$ relative to their minimum. 

To investigate the experimental strength functions further, they are compared to theoretical predictions. For the $E1$ part of the total $\gamma$-strength function, the Kadmenski{\u{\i}}, Markushev and Furman (KMF) model~\cite{kad} described by
\begin{equation} 
f_{E1}(E_\gamma)=\frac{1}{3\pi^2\hbar^2c^2} \frac{0.7\sigma_{E1}\Gamma_{E1}^2(E_\gamma^2+4\pi^2T^2)} {E_{E1}(E_\gamma^2-E_{E1}^2)^2}
\label{eq:E1}
\end{equation}
is applied. Here, $\sigma_{E1}$ is the cross section, $\Gamma_{E1}$ is the width, and $E_{E1}$ is the centroid of the giant electric dipole resonance (GEDR). The Lorentzian parameters are taken from~\cite{CDFE} (see Table~\ref{tab:tab3}). The nuclear temperature on the final state, introduced to ensure a nonvanishing GEDR for $E_{\gamma} \rightarrow 0$, is given by $T(E_{f}) = \sqrt{U_{f}/a}$. 

For $f_{M1}$, which is supposed to be governed by the spin-flip $M1$ resonance \cite{voin1}, the Lorentzian giant magnetic dipole resonance (GMDR)
\begin{equation}
f_{M1}(E_\gamma)=\frac{1}{3\pi^2\hbar^2c^2} \frac{\sigma_{M1}E_\gamma\Gamma_{M1}^2} {(E_\gamma^2-E_{M1}^2)^2+E_\gamma^2\Gamma_{M1}^2}
\label{eq:M1}
\end{equation}
is adopted. 

The contribution from $E2$ radiation to the total strength function is assumed to be very small. However, for the sake of completeness, the $E2$ isoscalar reconance described by
\begin{equation}
f_{E2}(E_\gamma)=\frac{1}{5\pi^2\hbar^2c^2 E_\gamma^2} \frac{\sigma_{E2}E_\gamma\Gamma_{E2}^2} {(E_\gamma^2-E_{E2}^2)^2+E_\gamma^2\Gamma_{E2}^2}
\label{eq:E2}
\end{equation}
is included in the total, theoretical strength function. 

In lack of any established theoretical prediction of the observed increase at low $\gamma$ energy, this phenomenon is modelled by a simple power law as
\begin{equation}
f_{\rm upbend}(E_\gamma)=\frac{1}{3\pi^2\hbar^2c^2}  A E_\gamma^{-b},
\label{eq:upbend}
\end{equation}
where $A$ and $b$ are fit parameters. 

\begin{figure}[htb]
\centering
\includegraphics[height=15cm]{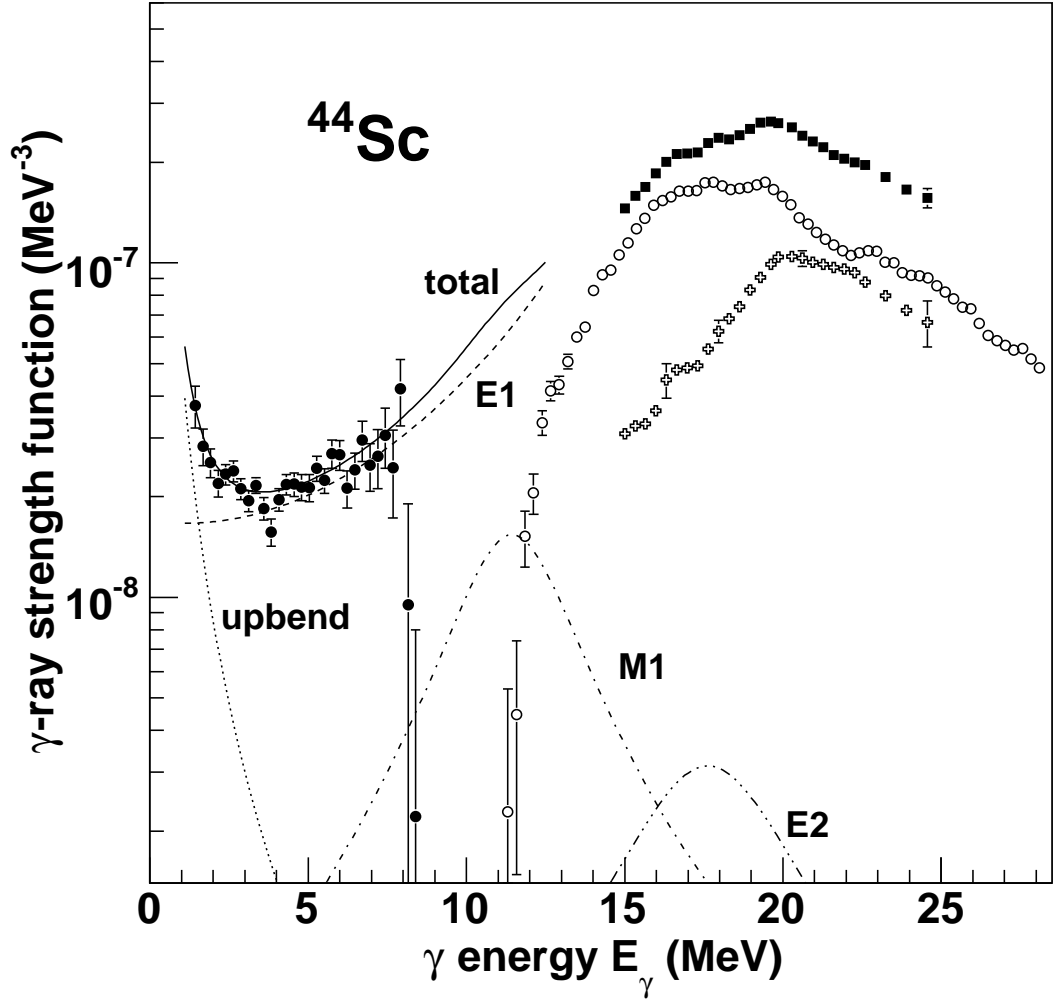}
\caption{The $\gamma$-strength functions of $^{44,45}$Sc from Oslo experiments (black dots) and GDR data from ($\gamma$,n) (white dots) and ($\gamma$,p) (white crosses) experiments~\cite{veyssiere,oikawa}. The black squares represent the summed strength from the ($\gamma$,n) and ($\gamma$,p) experiments for $E_{\gamma} = 15.0 - 24.6$ MeV. Also the total, theoretical strength function (solid line), the $E1$ tail from the KMF model (dashed line), the spin-flip $M1$ resonance (dashed-dotted line), the $E2$ isoscalar resonance (dashed-dotted line) and a fit to the upbend structure (dotted line) are shown.}
\label{fig:rsf_fitted}
\end{figure}

The total, theoretical $\gamma$-ray strength function is then given by
\begin{equation}
f_{\rm total} = {\kappa} (f_{E1} + f_{M1} + f_{\rm upbend}) + E_\gamma^{2} f_{E2},
\label{eq:totalrsf}
\end{equation}
where ${\kappa}$ is a renormalization factor that should be close to unity. All parameters employed are listed in Table~\ref{tab:tab3}, and the result for $^{44}$Sc is displayed in Fig.~\ref{fig:rsf_fitted}. It is seen that the theoretical strength function fits the data well. From Fig.~\ref{fig:rsf_fitted}, one would also conclude that the data points below $\sim 3$ MeV are not described by the standard models.

In Fig.~\ref{fig:rsf_fitted} also the photoneutron cross-section data from the reaction $^{45}$Sc($\gamma$,n)$^{44}$Sc~\cite{veyssiere} and the photoproton cross-section data from the reaction $^{45}$Sc($\gamma$,p)$^{44}$Ca~\cite{oikawa} are shown. The photoabsorbtion cross-section $\sigma (E_{\gamma})$ is converted into strength function through the relation
\begin{equation}
f (E_{\gamma}) = \frac{1}{3\pi^{2}\hbar^{2}c^{2}} \cdot \frac{\sigma (E_{\gamma})}{E_{\gamma}}.
\label{eq:trans}
\end{equation}
 
The ($\gamma$,n) and ($\gamma$,p) data exhaust $\sim 57$\% and $\sim 25$\% of the Thomas-Reiche-Kuhn sum rule, respectively~\cite{CDFE}. The summed strength of the two photoabsorption experiments for $E_{\gamma} = 15.0 - 24.6$ MeV is also displayed in Fig.~\ref{fig:rsf_fitted}, and it seems to fit reasonably well with the theoretical expectation and the Oslo data. Note that the photoabsorption cross-sections from the ($\gamma$,n) and ($\gamma$,p) reactions may have some overlap in strength in the energy region where the ($\gamma$,pn) channel is opened.  

\begin{figure}[htb]
\centering
\includegraphics[height=15cm]{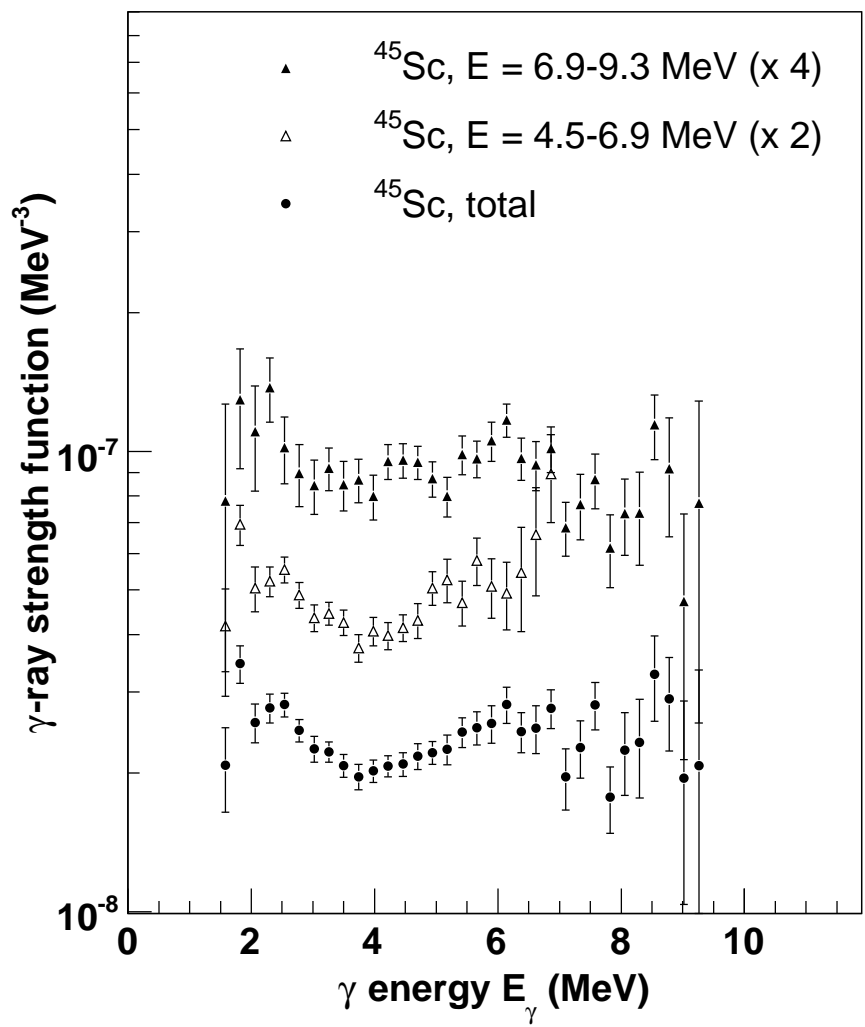}
\caption{The $\gamma$-strength function of $^{45}$Sc extracted from different excitation-energy regions together with the strength function obtained from the total excitation-energy region considered.}
\label{fig:rsf_test}
\end{figure}

For $\gamma$ energies below $\sim 3$ MeV, the $\gamma$-ray strength functions of $^{44,45}$Sc display an increase of a factor $\sim 3$ relative to their minimum. This behaviour has been observed in several medium-mass nuclei; first in $^{56,57}$Fe~\cite{voinov}, then recently in $^{93-98}$Mo~\cite{gutt7} and $^{50,51}$V~\cite{larsen}. For the iron and molybdenum isotopes, the upbend structure has been shown to be independent of excitation energy. This has also been tested for the Sc isotopes, as demonstrated in Fig.~\ref{fig:rsf_test}. Here, the $\gamma$-ray strength function of $^{45}$Sc has been extracted from two different excitation-energy regions (the intervals $4.5-6.9$ MeV and $6.9-9.3$ MeV), representing two independent sets of data. As seen in Fig.~\ref{fig:rsf_test}, the result is quite convincing. The general trends are very similar, and the enhancement at low $\gamma$ energies appears in both data sets. 

The physical origin of this low-energy enhancement in strength is not yet understood.  To check if the upbend feature could be due to peculiarities of the nuclear reactions or the Oslo method, a two-step cascade (n,$2\gamma$) experiment was carried out with $^{56}$Fe as a target~\cite{voinov}. This experiment confirmed the large increase in $\gamma$-ray strength observed in the Oslo data, but was unable to establish the character and multipolarity of the enhancement. To pin down the physical reason behind these observations, it is necessary to design and carry out experiments which have the possibility to determine the electromagnetic nature of this low-energy structure. Also, it would give better confidence to the findings to have independent confirmation of the increase from, e.g, (n,$2\gamma$) experiments on the Mo, V, and Sc isotopes as well.

\vspace{0.4cm}
\section{Summary and conclusions}

The nuclear level densities and the $\gamma$-ray strength functions of the scandium isotopes $^{44,45}$Sc have been measured from primary $\gamma$ rays using the Oslo method. The level densities display bump structures that cannot be obtained from standard statistical level-density models. A new, microscopic model to calculate the level density has been developed and applied on both nuclei, giving an overall good agreement with the experimental data. From the model, information on the average number of broken pairs and the parity asymmetry can also be extracted. 

The $\gamma$-ray strength functions are in general found to be increasing functions of $\gamma$ energy in the energy region examined in this work. The new data sets from the Oslo experiment are compared to theoretical models of the strength function and photoabsorbtion data, and the agreement seems to be good.  At low $\gamma$ energies a substantial enhancement of the total $\gamma$-ray strength is observed, that is not accounted for in any of the standard theories. As of today, this puzzling feature has no satisfying, physical explanation.   

\acknowledgments
Financial support from the Norwegian Research Council (NFR) is gratefully acknowledged. A.~Schiller acknowledges support from the U.~S. National Science Foundation, grant number PHY-06-06007.

\onecolumngrid
\newpage

\begin{table}
\caption{Parameters used for the back-shifted Fermi gas level density and the parameters from~\cite{egidy}.} 
\begin{tabular}{c|ccc|ccc|cccc|c}
\hline
\hline
	
Nucleus    & $E_1$    & $a$     &$\sigma$ & $E_{1}^{\dagger}$ & $a^{\dagger}$  &$\sigma^{\dagger}$& $B_p$  & $B_p + (\Delta E)/2$ & $D^{\ddagger}$ & $\rho$(proton res.) & $\eta$ \\ 
           & (MeV)   & (MeV$^{-1}$)  &  & (MeV)   & (MeV$^{-1}$) &    & (MeV)  &   (MeV)         & (eV)       & (MeV$^{-1}$)               \\
\hline

$^{44}$Sc  & -2.91 & 5.13 & 3.53 & -2.06   & 5.68  & 3.37  & 6.696  & 7.896  & 3243(324)  & 1855(392) & 1.12     \\
$^{45}$Sc  & -2.55 & 4.94 & 3.75 & -0.61   & 6.07  & 3.41  & 6.889  & 9.904  & 7874(496)  & 3701(760) & 1.26     \\
\hline
\hline
\end{tabular}
\\
$^{\dagger}$Calculated with the method of~\cite{egidy} \\
$^{\ddagger}$Calculated from proton-resonance data
\label{tab:tab1}
\end{table}

\begin{table}
\caption{Model parameters.} 
\begin{tabular}{c|c|cc|c|cc|cc}
\hline
\hline
Nucleus   &$\epsilon_2$&$\Delta_{\pi}$&$\Delta_{\nu}$& $A_{\rm rot}$ & $\hbar \omega _{0}$ & $\hbar \omega _{\rm vib}$ & $\lambda_{\pi}$ &  $\lambda_{\nu}$\\ 
          &            &      (MeV)   &     (MeV)    &      (MeV)    &  (MeV)  &       (MeV)  & (MeV) & (MeV)\\
\hline
$^{44}$Sc &     0.23   &  1.234       &  1.559       &    0.135      &   11.61 & 1.904       & 45.96 & 47.47  \\
$^{45}$Sc &     0.23   &  1.353       &  1.599       &    0.135      &   11.53 & 1.904       & 45.60 & 47.91    \\
\hline
\hline
\end{tabular}
\\
\label{tab:tab2}
\end{table}

\begin{table}
\caption{Parameters used for the theoretical $\gamma$-ray strength functions.} 
\begin{tabular}{c|ccc|ccc|ccc|ccc}
\hline
\hline
Nucleus   & ${\kappa}$ & $A$ & $b$    & $E_{E1}$ & $\sigma_{E1}$ & $\Gamma_{E1}$ &   $E_{M1}$ & $\sigma_{M1}$ & $\Gamma_{M1}$ & $E_{E2}$ & $\sigma_{E2}$ & $\Gamma_{E2}$\\ 
          &          &                &          & (MeV)      & (mb)            & (MeV)          &  (MeV)      &   (mb)          & (MeV)           & (MeV)      &   (mb)          & (MeV)           \\
\hline

$^{44}$Sc & 1.11(3)  & 0.52(10)        & 2.57(23) & 19.44      & 39.40           & 8.0            & 11.61 & 1.239 & 4.0 & 17.85 & 1.069 & 5.58 \\
$^{45}$Sc & 1.20(1)  & 1.62(9)        & 2.93(5)  & 19.44      & 39.40           & 8.0            & 11.53 & 1.214 & 4.0 & 17.71 & 1.047 & 5.57 \\
\hline
\hline
\end{tabular}
\\
\label{tab:tab3}
\end{table}

\newpage

\newpage

\newpage

\end{document}